\documentstyle[12pt]{article}

\begin{document}
\centerline{\bf Experimental scheme for quantum teleportation 
of a single-photon packet}
\vskip 1mm
\centerline{S.N.Molotkov}
\vskip 1mm
\centerline{\sl\small Institute of Solid State Physics 
       of Russian Academy of Sciences,}
\centerline{\sl\small Chernogolovka, Moscow distr., 142432 Russia}
\vskip 1mm
\begin{abstract}
Both complete protocol and optical setup for
experimental realization of quantum teleportation of unknown 
single-photon wave packet are proposed.
\end{abstract} 

E-mail: molotkov@issp.ac.ru
\vskip 1mm

Quantum mechanics prohibits cloning (copying) an unknown quantum state
(no cloning theorem [1]). Is it possible to transmit to a distant user
a previously unknown quantum state without sending that state itself?
Strictly speaking, any measurement aimed at obtaining classical information
to be sent to another observer changes the state itself without providing 
complete information about it. Production of a large number of identical copies
which can be measured many times to obtain complete information on their
common quantum state is prohibited by the no-cloning theorem. Thus, 
it is impossible to transmit information on a quantum state employing
only a classical communication channel. 

Quantum teleportation lifts this restriction if a quantum communication channel
is used in addition to the classical one. The idea of quantum teleportation
for the case of discrete quantum states (e.g. a spin-1/2 particle in 
an unknown state) was first proposed in Ref.[3]. A quantum channel is realized 
through the non-local EPR-correlations [2,3]\footnote{This term stems from the
well-known Einstein-Podolsky-Rosen effect [2]}. An EPR-pair is a pair
of particles described by an entangled state. Entanglement constitutes 
a special kind of quantum correlations which do not have
any classical analogies.

The quantum teleportation protocol described in Ref.[4] has the following
form. To teleport an unknown quantum state from user {\it A} to a 
distant user {\it B}, user {\it A} generates an EPR-pair. 
One of the particles of that EPR-pair remains with the user  
{\it A} while the second one is sent to the distant user {\it B}. 
User {\it A} performs a joint measurement over the particle in the 
unknown state to be teleported and his particle from the EPR pair thus
obtaining classical information. Because of the non-local 
correlations inherent in the EPR-pair, the measurement outcome
uniquely determines the resulting state of the second particle in
the EPR-pair sent to user {\it B}.
The state of the second particle coincides with the unknown state
to within a unitary rotation. Classical information obtained in the 
measurement is sent by user {\it A} to {\it B} and is used by the latter
to determine which unitary transformation should be performed to obtain
a new state identical to the original unknown state. In the course of 
teleportation user {\it A} obtains no information on the teleported
unknown state.

Quantum teleportation has recently been demonstrated experimentally
for a photon in an unknown polarization state [5,6].

The problem of teleportation of the wave function of a particle
in the one-dimensional case where position and momentum are the
two continuous dynamic variables was investigated in Ref.[7]
using the wave function proposed in Ref.[2] to describe an EPR pair.
Studied in a recent work [8] was the quantum teleportation of 
a quantum state described by the dynamical variables 
(the unknown state in Ref.[8] corresponds to a single-mode photon state)
for the case of non-ideal EPR-correlations. A quadrature-squeezed
state was used as an EPR-state while the measurement procedure
actually corresponded to the homodyne detection.

Proposed in the present paper is a new scheme (a complete protocol
and its experimental realization) for the teleportation of a 
multi-mode state, i.e. a single-photon wave packet, employing an EPR-pair
in an entangled state with respect to the energy--time variables.
 
To simplify further formulas, we shall assume that the packet polarization
state is known. The discussion below is also applicable to the
case of unknown polarization which can be accounted for introducing an 
additional subscript. The state of a single-photon wave packet can be 
written as (see, e.g. Ref.[9]) 
\begin{equation}
|1\rangle_3 = 
\int_{0}^{\infty}d\omega f(\omega)\hat{a}^{+}(\omega)|0\rangle =
\int_{0}^{\infty}d\omega f(\omega)|\omega\rangle_3,
\end{equation}
\begin{displaymath}
[\hat{a}(\omega),\hat{a}^+(\omega')]=I\delta(\omega-\omega'),\quad
\int_{0}^{\infty} |f(\omega)|^2  d\omega =1,
\end{displaymath}
where $\hat{a}^+(\omega)$, $\hat{a}(\omega)$ are the creation and
annihilation operators of a single-mode Fock state $|\omega\rangle_3$, 
$|0\rangle$ is the vacuum state, $f(\omega)$ is the packet amplitude 
Subscript 3 labels the channel number (see Fig.1). 
The density matrix at an arbitrary time is
\begin{equation}
\rho(3)=
\left(\int_{0}^{\infty}d\omega e^{-i\omega t} f(\omega)|\omega\rangle_3\right)
\left(\int_{0}^{\infty}d\omega'{}_3\langle \omega| e^{i\omega' t}f^*(\omega')
\right) 
\end{equation}
In our case the state of an EPR-pair can be written as 
(these photon EPR-pairs are produced in the parametric energy down-conversion
processes [10]) 
\begin{equation}
|\psi_{EPR}\rangle_{1,2} = 
\int_{0}^{\infty}d\omega  
|\omega\rangle_1\otimes|\Omega-\omega\rangle_2,\quad 
\rho_{_{EPR}}(1,2)=|\psi_{EPR}\rangle_{1,2}\mbox{ } {}_{1,2}\langle \psi_{EPR}|,
\end{equation}
where $\Omega$ is the pumping frequency and 1, 2 are the channel numbers (Fig.1).
The normalization of the state (3) is insignificant for further analysis.

According to the general scheme [11--13], quantum mechanical 
measurements are described by positive operators realizing the identity
resolution. Measurements of the variables corresponding to
self-adjoint operators are associated with the orthogonal resolutions.
Parameters (such as time and rotation angle) are not associated with any
self-adjoint operators so that the corresponding measurements are described
by the non-orthogonal identity resolution [11-13].

In the present paper the basic idea of proposed teleportation scheme
consists in the usage of a joint (entangled) time--energy measurement
performed on a pair of photons one of which belongs to the EPR-pair
and the second photon in the unknown state to be teleported. 
The measurement is given by the non-orthogonal identity resolution [14]
\begin{equation}
\int\int M(dtd\Omega_{+})=
\int\int R^{+}R\mbox{ }(dtd\Omega_{+})=I,
\end{equation}
where $R$ is the ``reduction'' operator, $M(dtd\Omega_{+})$ describes 
a quantum operation and is a positive operator valued measure, 
POVM, (the details can be found, e.g. in Refs.[15,16]); we have [14]
\begin{equation}
M(dtd\Omega_{+})=
\end{equation}
\begin{displaymath}
\left(\int d\omega_{-} e^{i\omega_{-}t}
|\omega_{+}+\omega_{-}\rangle_1
\otimes|\omega_{+}-\omega_{-}\rangle_3\right)
\left(\int d\omega_{-}'e^{-i\omega_{-}'t}
{}_3\langle\omega_{+}-\omega_{-}'|
\otimes{}_1\langle\omega_{+}+\omega_{-}'|\right)
\frac{\textstyle dtd\omega_{+}}{\textstyle 2\pi},
\end{displaymath}
$\omega_{\pm}=\frac{\Omega_{\pm}}{2}$.
The integration covers the frequency ranges corresponding
to positive arguments of the Fock states. It should be emphasized
that the frequency $\omega_{+}$ is common to bra- and ket-states. 

According to the general concepts of the quantum measurement theory
[10--13,15,16], application of a quantum operation
(measurement) to a system described by the density matrix $\rho$
transforms it to a new state 
\begin{displaymath}
\rho\rightarrow 
\frac{\textstyle  R_i\rho R^{+}_{i} }
{ \textstyle \mbox{Tr} \{ R_i\rho R^{+}_{i} \} }.
\end{displaymath}
The probability of the $i$-th outcome is given by the formula
$\mbox{Pr}=\mbox{Tr} \{ R_i\rho R^{+}_{i} \}$, where $E_i=R^{+}_{i}R_i$ 
is the POVM element.

In our case, after the measurement performed by user {\it A},
the state of the second photon from the EPR-pair observed by user {\it B}
is given by the density matrix
\begin{equation}
\tilde{\rho}(2)=
\frac{\textstyle \mbox{Tr}_{1,3} 
\{ \rho_{_{EPR}}(1,2)\otimes\rho(3)M(dt d\Omega_{+}) \} }
{ \textstyle \mbox{Pr}\{dt d\Omega_{+}\} },
\end{equation}
\begin{equation}
\mbox{Pr}\{dt d\Omega_{+}\}=
\mbox{Tr}_{1,2,3} \{ \rho_{_{EPR}}(1,2)\otimes\rho(3)M(dt d\Omega_{+}) \}, 
\end{equation}
\begin{equation}
\tilde{\rho}(2)=
\left(\int d\omega e^{-i(\Omega_{+}/2+\omega) t} 
f(\Omega_{+}-\omega)|\Omega-\omega\rangle_2\right)
\left(\int d\omega'{}e^{i(\Omega_{+}/2+\omega') t}
{}_{2}\langle\Omega- \omega'| f^*(\Omega_{+} - \omega') 
\right) 
\end{equation}
Formally, the measurement (4) corresponds to the situation where
the measurement moment $t$ and frequency $\Omega_{+}$ are chosen
by the experimentator, and the positive result probability is given by Eq. (7).
The teleportation will be ideal if the chosen  $\Omega_{+}=\Omega$ 
(detector registration frequency $\omega_{+}$ 
coincides with the pumping frequency $\Omega$). In that case it follows from
Eq. (6) that the state
$\tilde{\rho}(2)$ coincides with $\rho(3)$ to within a phase factor
which can be eliminated by user {\it B} if user {\it A}
sends to him the registration time $t$ via a classical channel.
Note that the registration time $t$ does not depend on unknown state
$\rho(3)$ if $\Omega_{+}=\Omega$. 

Physically, the measurement (4) can be understood in the following way. 
User {\it A} has a continuum of detectors ``tuned'' to the frequencies in the
($0,\infty$) range, and each of them can fire at an arbitrary time
$t$, formally in the infinite ($-\infty,\infty$) interval. 
The probability for the detector tuned to frequency $\Omega_+$ to fire
at time $t$ is given by Eq. (7). The firing probability does not depend on 
time only for the detector tuned to frequency $\Omega$. The teleportation 
will only be ideal if the detector tuned to $\Omega$ fires. In that case
it follows from Eq. (7) that the firing probability does not depend on
the unknown input state. User {\it A} does not obtain any information
on the teleported state.

Since the measurements can be performed at spatially separated points,
all the times in the above formulas should be understood as the reduced
time-of-flight corrected times ($t\rightarrow t-x/c$). It will be seen below
that this point is insignificant in the case of ideal teleportation.

The major problem in realizing the single-photon packet teleportation
is the implementation of the measurement given by Eq. (4). The measurement (4)
on a pair of photons is an intermediate case between the time and energy 
measurements; its experimental realization is described below. The idea is 
to convert a photon pair into a single photon which then is measured by 
narrow-band photodetector. The latter can be easily realized experimentally.

The experimental setup is presented in Fig.1. 
The first non-linear crystal with the second-order susceptibility 
$\chi$ and a narrow-band filter tuned to the frequency 
$\Omega$ serve to generate the EPR-pair in channels 1 and 2. 
The channel 3 is used to feed the unknown single-photon packet. 
The latter can be prepared by exciting a two-level system with
a $\pi$-pulse a long time ago. The second non-linear crystal, 
the narrow-band filter behind it tuned to frequency $\Omega$, and
then a standard photodetector realize the measurement (4). 
The teleported state arises in channel 2.

Consider step by step the input state evolution in the optical scheme.
After the first narrow-band filter in front of the first non-linear crystal
the state is described by a monochromatic state with the density matrix
\begin{equation}
\rho_{in}(in)=|\Omega\rangle_{in}\mbox{ }{}_{in}\langle\Omega|,
\end{equation}
which can be obtained by cutting a narrow band by the first filter from
an auxiliary single-photon packet fed into the input channel $in$ (Fig.1). 
The photon-photon interaction in the nonlinear crystal is described in the 
interaction representation by the following Hamiltonian
(details can be found in Refs. [17,18])
\begin{equation}
H_1(t)=\chi\int d{\bf x}E_{in}^{(+)}({\bf x},t)E_{1}^{(-)}({\bf x},t)
E_{2}^{(-)}({\bf x},t) +h.c.,
\end{equation}
all the insignificant constants are assumed to be included in the
definition of $\chi$, which as usually [17,18] will be assumed to be 
frequency-independent. It is convenient to present the electric field operators 
in the form [9] 
\begin{equation}
E_{i}^{(-)}({\bf x},t)=
\frac{1}{\sqrt{2\pi}}
\int_{0}^{\infty}d\omega e^{i(\omega t-{\bf kx})}
\hat{a}^{+}(\omega)|0\rangle_{i}=
\frac{1}{\sqrt{2\pi}}
\int_{0}^{\infty}d\omega e^{i(\omega t-{\bf kx})}|\omega\rangle_{i},
\end{equation}
where $i$ is the channel number. Analogously for $E_{i}^{(+)}({\bf x},t)$. 
Taking into account Eq. (11) we have
\begin{equation}
H_1(t)=\frac{\chi}{(2\pi)^{3/2}}
\int\int\int
d\omega_1d\omega_2 d\omega_{in}
e^{ it( \omega_1+\omega_2-\omega_{in} ) }
|\omega_1\rangle_1\otimes |\omega_2\rangle_2\mbox{ }{}_{in}\langle\omega_{in}|
\int_{vol}d{\bf x} e^{-i{\bf x}({\bf k_1+k_2-k_{in}})}+h.c.
\end{equation}
In the second integral the integration is performed throughout the entire 
crystal yielding the 
$\delta$-symbol with respect to momentum, and results in the
phase synchronism conditions 
[18] (${\bf k_1+k_2}={\bf k_{in}}$), which will be assumed satisfied (below 
it means that ${\bf k_2}\|{\bf k_3}$, Fig.1). In the subsequent
formulas the quantity $\chi$ is understood as the renormalized constant
corrected for the additional factors arising from the second integral.
The first-order susceptibility which is always present can be neglected 
in our analysis since the corresponding terms in the Hamiltonian
do not contribute to the {\it out} channel. 

The state after the first crystal in channels 1 and 2 is described by 
the density matrix
\begin{equation}
\rho_{EPR}(1,2)=S(t)\rho_{in}(in)S^{-1}(t),
\end{equation}
where $S(t)$ is the $S$-matrix
\begin{equation}
S(t)=e^{i\int_{-\infty}^{t}H_1(t')dt'}=1+S^{(1)}+S^{(2)}+\ldots.
\end{equation}
In the first order with respect to $\chi$ one has
\begin{equation}
S^{(1)}=i\chi\int\int
d\omega_1 d\omega_{in} |\omega_1\rangle_1\otimes 
|\omega_{in}-\omega_1\rangle_2 
\mbox{ }{}_{in}\langle\omega_{in}| +h.c.  
\end{equation}
The upper integration limit in the exponent in $S$ can be replaced by 
$\infty$ which is physically actually related to the fact that the input 
state is a monochromatic one (roughly speaking infinitely extended in time) 
and the teleportation process is formally stationary. 
To within an unimportant normalization constant, the state in channels  
1 and 2 is described by the density matrix 
\begin{equation}
\rho_{EPR}(1,2)=\chi^2
\left(\int_{0}^{\infty}d\omega|\omega\rangle_1\otimes|\Omega-\omega\rangle_2
\right)
\left(\int_{0}^{\infty}d\omega'{}_{1}\langle\omega'|\otimes
{}_{2}\langle\Omega-\omega'|\right).
\end{equation}
Measurement by a photodetector which has a narrow-band filter tuned to the
frequency $\Omega$ installed in front of it is formally described by the 
projector 
$P(\Omega)=|\Omega\rangle_{out}\mbox{ }{}_{out}\langle\Omega|$.

The teleported state in channel 2 after the registration by the photodetector 
is described by the density matrix (again to within the normalization constant) 
\begin{equation}
\tilde{\rho}(2)=\mbox{Tr}_{out}\left\{
S(t)\rho_{in}(in)\otimes\rho(3)S^{-1}(t)P(\Omega)\right\},   
\end{equation}
where $S(t)$ is now the full $S$-matrix of the entire optical scheme,
\begin{equation}
S(t)=e^{i\int_{-\infty}^{t}[H_1(t') + H_2(t')]dt'}=1+S^{(1)}+S^{(2)}+\ldots,
\end{equation}
where $H(t)_2$ is th second non-linear crystal Hamiltonian which coincides 
to within the subscripts interchange with $H(t)_1$ in Eq. (12).
Contributing to the teleportation processes are the $S$-matrix 
terms of the form
\begin{equation}
S^{(2)}\propto
\end{equation}
\begin{displaymath}
\chi^2
\left(\int\int d\omega_1 
d\omega_{in} |\omega_1\rangle_1 \otimes|\omega_{in}-\omega_1\rangle_2  
\mbox{ }{}_{in}\langle\omega_{in}|\right)
\left(\int\int d\omega'_1 d\omega_{out} 
|\omega_{out}\rangle_{out}\mbox{ } {}_{1}\langle\omega'_{3}|\otimes
{}_{in}\langle\omega_{out}-\omega'_{1}|\right). 
\end{displaymath}
Taking into account Eq. (19), the density matrix in channel 2 to within
the normalization constant coincides with the initial density matrix of 
he unknown wave packet
\begin{equation}
\tilde{\rho}(2)=\chi^4
\left(\int_{0}^{\infty}d\omega f(\omega)|\omega\rangle_3\right)
\left(\int_{0}^{\infty}d\omega'{}_3\langle \omega| f^*(\omega')
\right) 
\end{equation}
It also follows from Eq. (17) that the detection probability in the
output channel ($out$) does not depend on the unknown state and 
is proportional to
\begin{equation}
\mbox{Pr}=\mbox{Tr}_{2,out}\left\{
S(t)\rho_{in}(in)\otimes\rho(3)S^{-1}(t)P(\Omega)\right\}
\propto \chi^4.   
\end{equation}
In the outlined scheme the classical channel is used to inform the distant
user of the fact that a photodetector fired, and in that case the
teleportation is assumed to be successful. The probability (efficiency)
of the teleportation process is small to the measure of $\chi^4$.
The fraction of false photodetection firings when a wrong state will be
teleported due to the terms of higher orders in $\chi$ in the $S$-matrix
has an additional smallness in $\chi^2$. Note that formally the teleportation 
process is stationary (requires infinite time), since a monochromatic
state should be prepared. In that case the probability of the
photodetector firing in the $out$ channel does not depend on time
and does not depend on the input state. In that case user {\it A}
obviously acquires zero information on the teleported state.

Of course, quantum teleportation does not allow information transmission
faster than the speed of light. In the present scheme an intuitive and
qualitative explanation is as follows. Since the input state is monochromatic 
and always intuitively assumed to be non-localized (infinitely extended), 
the latter means that the field is ``prepared in advance'' throughout the
entire space, including the locations of both distant users 
{\it A} and {\it B}. The measurement performed by user {\it A} transforms
the entire system to a new state---reduces the state vector 
``immediately'' and ``everywhere'' for the entire system. 
This assumption is usually considered as counterintuitive. 
However, these ``immediately'' and ``everywhere'' do not result in 
faster than light communication. To transmit classical information 
from {\it A} to {\it B} with the help of a teleported state,
one requires a classical channel from {\it A} to {\it B} 
to tell that the detector fired and the teleportation was successfully
completed. Classical communication channel assumes sending a classical
object from {\it A} to {\it B} whose velocity cannot exceed that of light.
The problem of the field ``prepared in advance'' everywhere is closely
related to the photon localizability (to be more precise, non-localizability)
(e.g. see Refs.[19--23]). As far as I know this problem has not yet been
discussed in detail in the context of quantum teleportation.

It should be noted that the teleportation process can be reformulated
in the Feynman diagram language, the averaging being performed
over the stationary state
$|\Omega\rangle\otimes \int_{0}^{\infty}d\omega f(\omega)|\omega\rangle$ 
corresponding to the input monochromatic state and the single-photon state.
In the case of ideal teleportation this state is also the output state.
In that sense the process is stationary and the averaging is performed over
the stationary state which should not necessarily be the ground state.
In that case the diagrammatic technique is developed in the same way as 
in the Keldysh method [24].
 
The author is grateful to B.A.Volkov, M.V.Lebedev, 
S.S.Nazin, and S.T.Pavlov for discussions.
The work was supported by the Russian Fund for Basic Research
(Grant \# 96-02-19396), and the Program ``Advanced Technologies in
Micro- and Nanoelectronics'' (Grant \# 02.04.329.895.3).


\begin{thebibliography}{99}
\bibitem{1}W.K.Wootters, W.H.Zurek, Nature, {\bf 299}, 802 (1982).
\bibitem{2}A.Einstein, B.Podolsky, N.Rosen, Phys. Rev., {\bf 47}, 777 (1935).
\bibitem{3}J.S.Bell, {\it Speakable and Unspeakable in Quantum 
Mechanics}, (Cambridge University Press, Cambridge, England, 1988).
\bibitem{4}C.H.Bennett, G.Brassard, C.Crepeau, R.Jozsa, A.Peres, 
W.K.Wootters, Phys. Rev. Lett., {\bf 70}, 1895 (1993).
\bibitem{5}B.Boumeester, Jian-Wei Pan, K.Mattle, M.Eibl, H.Weinfurter, 
A.Zeilinger, Nature, {\bf 390}, 575 (1997).
\bibitem{6}D.Boschi, S.Branca, F.De Martini, L.Hardy, S.Popescu, 
Phys. Rev. Lett., {\bf 80}, 1121 (1998).
\bibitem{7}L.Vaidman, Phys. Rev., {\bf A49}, 1473 (1994).
\bibitem{8}S.Braunstein, H.J.Kimble, Phys. Rev. Lett., {\bf 80}, 869 (1998).
\bibitem{9}R.A.Campos, B.E.A.Saleh, M.C.Teich, Phys. Rev., {\bf A42}, 4127 (1990).
\bibitem{10}P.G.Kwait, K.Mattle, H.Weinfurter, A.Zeilinger, A.V.Sergienko,
Y.H.Shin, Phys. Rev. Lett., {\bf 75}, 4337 (1995).
\bibitem{11}A.S.Holevo, {\it Probabilistic and Statistical Aspects
of Quantum Theory}, ``Nauka'', M. 1980 (in Russian).
\bibitem{12}P.Busch, M.Grabowski, P.J.Lahti, {\it Operational Quantum
Physics}, Springer Lecture Notes in Physics, {\bf v. 31}, 1995.
\bibitem{13}K.Krauss, {\it States, Effects and Operations}, Springer-Verlag, 
Berlin, 1983.
\bibitem{14}S.N.Molotkov, in {\it LANL E-Print Archive, 
http:// xxx.lanl.gov/ quant-ph/9805045}.
\bibitem{15}B.W.Schumacher, Phys. Rev., {\bf A54}, 2614 (1996).
\bibitem{16}M.A.Nielsen, C.M.Caves, Phys. Rev., {\bf A55}, 2547 (1997).
\bibitem{17}C.K.Hong, L.Mandel, Phys. Rev., {\bf A31}, 2409 (1985).
\bibitem{18}P.W.Milonni, H.Fearn, A.Zeilinger, Phys. Rev., 
{\bf A53}, 4556 (1996).
\bibitem{19}T.D.Newton, E.P.Wigner, Rev. Mod. Phys., {\bf 21}, 400 (1949).
\bibitem{20}A.S.Wightman, Rev. Mod. Phys., {\bf 34}, 845 (1962).
\bibitem{21}K.Kraus, {\it Position observable of photon, in: \it
The Uncertainty Principle and Foundations of Quantum Mechanics}, 
Eds. W.C.Price and S.S.Chissik, John Wiley\& Sons, New York, p.293, (1976).
\bibitem{22}H.Bacry, {\it Localizability and Space in Quantum Physics, in:
Lecture Notes in Physics}, Eds. H.Araki et al, Springer-Verlag, Vol.308, (1988).
\bibitem{23}J.A.Brooke, F.E.Schroek, J. of Math. Phys., {\bf 37}, 5958 (1996).
\bibitem{24}L.V.Keldysh, ZhETF, 47 (1964) 1515 (Sov. JETP, 20 (1965) 1018).
\end{thebibliography}
\end{document}